# Skellam Rank: Fair Learning to Rank Algorithm Based on Poisson Process and Skellam Distribution for Recommender Systems

Hao Wang
*Ratidar Technologies LLC*
Beijing, China
haow85@live.com

*Abstract*—Recommender system is a widely adopted technology in a diversified class of product lines. Modern day recommender system approaches include matrix factorization, learning to rank and deep learning paradigms, etc. Unlike many other approaches, learning to rank builds recommendation results based on maximization of the probability of ranking orders. There are intrinsic issues related to recommender systems such as selection bias, exposure bias and popularity bias. In this paper, we propose a fair recommender system algorithm that uses Poisson process and Skellam distribution. We demonstrate in our experiments that our algorithm is competitive in accuracy metrics and far more superior than other modern algorithms in fairness metrics.

## I. INTRODUCTION

Recommender system is one of the most successful data mining technologies nowadays. It was introduced in late 1980's and 1990's and quickly developed into a major research field in artificial intelligence area. Unlike other artificial intelligence technologies such as natural language processing and computer vision, recommender system became commercializable fairly quickly. Important application scenarios of recommender systems include E-commerce websites, cultural record websites, music streaming services, social networks, and video social networks.

Recommender system is so successful because it saves billions of US Dollars on marketing expenses globally for the product owners each year, and plus, it entices a significant volume of sales and traffic for growing and grown-up corporations. Report claims 30% - 40% of Amazon sales could be attributed to the company's recommender system product lines. The statistical figure is a lion share of the company's total annual revenues.

The most fundamental idea of recommendation algorithm design is to come up with a loss function that minimizes the difference between predicted preference values of users and the true preference values of users. The metrics used in this sense are scores such as MAE (mean absolute error) and RMSE (rooted mean squared error). Main school algorithms in this area include collaborative filtering approaches and matrix factorization techniques.

Around the year 2010, a different school of thoughts were introduced into the area. Instead of minimizing the accuracy metric, researchers proposed to use ranking metrics to measure the performance of recommendation algorithms. The framework is termed Learning to Rank, and comprises of 3 different variants - Pointwise Learning to Rank, Pairwise Learning to Rank [1][2], and Listwise Learning to Rank [3][4]. One of the most famous Learning to Rank algorithm is Bayesian Personalized Ranking [1], which has been wide spread in both academia and industry.

The learning to rank algorithms are typically used as a second phase ranking procedure following the first stage of candidate generation procedure. The goal of the hybrid model is to enhance the overall performance of the system using combined power of different models.

Since the year of 2017, fairness becomes a new hot topic of the artificial intelligence field [5][6][7]. Although we have achieved enough accurate algorithms that could satisfy the need of product owners, we are yet to solve many intrinsic problems of recommender systems. Fairness is one of such problems, and it refers to a broad range of issues such as demographic unfairness and bias problems including selection bias [8], exposure bias [9], popularity bias [10], etc.

In this paper, we mainly focus on solving popularity bias issues of the recommender systems. One of the most commonly used approaches is regularization, namely by penalizing the loss function of learning to rank, and this technique is not only limited to learning to rank, but is also applicable in other paradigms such as matrix factorization [11] [12].

We adopt a different approach in this paper, we discover that by formulating the user item rating behavior as a Poisson process, we are able to model the pairwise ranking problem of recommender systems using Skellam distribution. As the end result of our mathematical modeling process, we are able to obtain a learning to rank algorithm that is competitively accurate but far more superior than other modern algorithms when evaluated with fairness metrics.

## II. RELATED WORK

Recommender systems incorporates a very broad spectrum of techniques and approaches such as collaborative filtering, matrix factorization, learning to rank and deep learning approaches. The product-wise classification is also diversified, we have context-aware recommendation [13][14][15], session-based recommendation [16][17][18], group recommendation [19][20][21], etc.

One of the most popular learning to rank approaches is Bayesian Personalized Ranking [1], which maximizes the probability of pairwise ranking orders. On the contrary, Collaborative Filtering [22] and Matrix Factorization paradigms [23][24] focus on optimizing the accuracy metrics such as Mean Absolute Error (MAE) and Rooted Mean Squared Error (RMSE). Another popular learning to rank technique is Collaborative Less is More Filtering [3], which adopts a list-wise approach for learning to rank problems.

Fairness issues in recommender systems have aroused deep interest in the research community. E. Chi [25] proposed a matrix factorization variant named focused learning that penalize the matrix factorization loss function with a regularization term. H. Wang introduced Zipf Matrix Factorization [5] and KL-Mat [6] following the same school of thoughts with different types of regularizers. Since 2020, major research venues such as SIGIR and WWW have witnessed an outburst of fair learning to rank algorithms [26][27][28].

In addition to the regularization framework, other researchers adopt a different perspective to model the problem. Researchers such as H. Wang build fair models with modifications to the underlying theories such as MatRec [29] and Pareto Pairwise Ranking [30].

Poisson process has been introduced into the field in as early as 2015 [31][32]. The basic idea is to model the user item rating behavior using Poisson process and build the statistical model into the framework of topic models.

## III. SKELLAM DISTRIBUTION

Poisson distribution is a widely adopted distribution used to model the occurrence of the rare event. The formal PMF definition of Poisson distribution is as follows :

$$P(X = k) = \frac{\lambda^k e^{-\lambda}}{k!}$$

where k represents the number of occurrences of rare events.

The mean and variance of the random variable X is equal to $\lambda$ :

$$\lambda = E[X] = Var[X]$$

The difference of 2 Poisson variables is a Skellam variable, which is defined as follows :

$$P(X_1 - X_2 = k) = e^{-(\lambda_1+\lambda_2)}\left(\frac{\lambda_1}{\lambda_2}\right)^{k/2} I_k\left(2\sqrt{\lambda_1\lambda_2}\right)$$

$I_k$ denotes the Bessel Function of the First Kind, and can be approximated in the following way :

$$I_k(X) = \sum_{m=0}^{\infty} \frac{(-1)^m}{m!\,(m+k)!}\left(\frac{X}{2}\right)^{2m+k}$$

Researchers have been modeling the user item rating behavior with Poisson process, namely :

$$R_{i,j} = \frac{\lambda^{R_{i,j}} e^{-\lambda}}{R_{i,j}!}$$

Since the parameter $\lambda$ is equivalent to the expected values of the Poisson variable, the above formula can be rewritten as follows :

$$R_{i,j} = \frac{\left(\frac{1}{n}\sum_{j=1}^{n} R_{i,j}\right)^{R_{i,j}} e^{-\left(\frac{1}{n}\sum_{j=1}^{n} R_{i,j}\right)}}{R_{i,j}!}$$

In recommender systems' algorithmic design, we can replace $R_{i,j}$ with $U_i^T \cdot V_j$ , as defined in the matrix factorization framework, and we acquire the following formula for $R_{i,j}$:

$$\widetilde{R_{i,j}} = \frac{\left(\frac{1}{n}\sum_{j=1}^{n} U_i^T \cdot V_j\right)^{U_i^T \cdot V_j} e^{-\left(\frac{1}{n}\sum_{j=1}^{n} U_i^T \cdot V_j\right)}}{U_i^T \cdot V_j!}$$

Plug in the formula into the loss function of matrix factorization framework, we obtain the following loss function :

$$L = \sum_{i=1}^{n}\sum_{j=1}^{m} \left(\widetilde{R_{i,j}} - R_{i,j}\right)^2$$

This is the fundamental idea behind the PoissonMat algorithm. We discussed the basics of the algorithm to facilitate our new invention in this paper in the next section. Our newly invented algorithm (which we call Skellam Rank) also uses the idea of Poisson process in its underlying theory.

## IV. SKELLAM RANK

We adopt a pairwise learning to rank approach to model the recommender system problem, i.e. :

$$L = \sum_{i=1}^{n}\sum_{w=1}^{n}\sum_{j=1}^{m}\sum_{k=1}^{m} P(R_{i,j} > R_{w,k}) I(R_{i,j} > R_{i,k})$$

We model the user item rating behavior as Poisson distribution, and therefore pairwise order probability becomes Skellam distribution, as defined in the following way :

$$P(R_{i,j} - R_{w,k}) = e^{-(E_i+E_w)}\left(\frac{E_i}{E_w}\right)^{\frac{E_i-E_w}{2}} I_k\left(2\sqrt{\lambda_1\lambda_2}\right)$$

where $I_k$ denotes the Bessel Function of the First Kind :

$$I_k = \sum_{t=0}^{\infty} \frac{(-1)^t}{t!(t+E_i-E_w)!}\left(\sqrt{E_iE_w}\right)^{2t+E_i-E_w}$$

and :

$$E_i = \frac{1}{n}\sum_{j=1}^{n}\frac{R_{i,j}}{n}, \quad E_w = \frac{1}{n}\sum_{j=1}^{n}\frac{R_{w,j}}{n}$$

After plugging in the Skellam distribution

$$L = \sum_{i=1}^{n}\sum_{w=1}^{n}\sum_{j=1}^{m}\sum_{k=1}^{m} e^{-(E_i+E_w)}\left(\frac{E_i}{E_w}\right)^{\frac{E_i-E_w}{2}} I_k\left(2\sqrt{E_iE_w}\right)$$

We use matrix factorization results to denote $R_{i,j}$ :

$$R_{i,j} = U_i^T \cdot V_j$$

To simplify the optimization procedure, we set $I_k$ to be irrelevant to U and V while preserving the values of $R_{i,j}$. If we apply the Stochastic Gradient Descent technique to optimize for the U and V in the loss function, the loss function L can be further simplified in the following way :

$$L = I_k e^{-(U_i^T \cdot V_j + U_w^T \cdot V_k)}\left(\frac{U_i^T \cdot V_j}{U_w^T \cdot V_k}\right)^{\frac{U_i^T \cdot V_j - U_w^T \cdot V_k}{2}}$$

Computing the update rules for the loss function , we have :

$$\frac{\partial L}{\partial U_i} = \frac{I_k t_2 t_4 t_5 t_0^{t_3-1}}{2}V_j + \frac{I_k t_4 t_5 t_6 log(t_0)}{2}V_j - \frac{I_k t_4 t_5 t_6 log(t_1)}{2}V_j - I_k t_4 t_5 t_6 V_j$$

where :

$t_0 = U_i^T \cdot V_j$, $t_1 = U_w^T \cdot V_k$, $t_2 = t_0 - t_1$, $t_3 = \frac{t_2}{2}$, $t_4 = t_1^{-t_3}$, $t_5 = exp(-(t_0+t_1))$, $t_6 = t_0^{t_3}$

The partial derivative with respect to $U_w$ is :

$$\frac{\partial L}{\partial U_w} = \frac{I_k t_4 t_5 t_6 log(t_1)}{2}V_k - \frac{I_k t_4 t_7 log(t_0)}{2}V_k - \frac{I_k t_2 t_7 t_1^{-(1+t_3)}}{2}V_k - I_k t_4 t_5 t_6 V_k$$

where :

$t_0 = U_i^T \cdot V_j$, $t_1 = U_w^T \cdot V_k$, $t_2 = t_0 - t_1$, $t_3 = \frac{t_2}{2}$, $t_4 = t_1^{-t_3}$, $t_5 = t_0^{t_3}$, $t_6 = exp(-(t_0+t_1))$, $t_7 = t_5 t_6$

The partial derivative with respect to $V_j$ is :

$$\frac{\partial L}{\partial V_j} = \frac{I_k t_2 t_4 t_5 t_0^{t_3-1}}{2}U_i + \frac{I_k t_4 t_5 t_6 log(t_0)}{2}U_i - \frac{I_k t_4 t_5 t_6 log(t_1)}{2}U_i - I_k t_4 t_5 t_6 U_i$$

where :

$t_0 = U_i^T \cdot V_j$, $t_1 = U_w^T \cdot V_k$, $t_2 = t_0 - t_1$, $t_3 = \frac{t_2}{2}$, $t_4 = t_1^{-t_3}$, $t_5 = exp(-(t_0+t_1))$, $t_6 = t_0^{t_3}$

The partial derivative with respect to $V_k$ is :

$$\frac{\partial L}{\partial V_k} = \frac{I_k t_4 t_5 t_6 log(t_1)}{2}U_w - \frac{I_k t_4 t_7 log(t_0)}{2}U_w + \frac{I_k t_2 t_7 t_1^{-(1+t_3)}}{2}U_w - I_k t_4 t_5 t_6 U_w$$

where :

$t_0 = U_i^T \cdot V_j$, $t_1 = U_w^T \cdot V_k$, $t_2 = t_0 - t_1$, $t_3 = \frac{t_2}{2}$, $t_4 = t_1^{-t_3}$, $t_5 = t_0^{t_3}$, $t_6 = exp(-(t_0+t_1))$, $t_7 = t_5 t_6$

After acquisition of the partial derivatives of the parameters, we are able to compute for the optimal values of parameters using Stochastic Gradient Descent (SGD) rules. In the next section, we will demonstrate that Skellam Rank is competitive with modern day recommender system algorithms, and it achieves the best score when evaluated by fairness metric.

The Pseudo-code for Skellam Rank is as follows :

```
Function Skellam-Rank :

1    Read input data into the user item rating matrix R
2    For iter in 1: max_iter_number:
3        User_sample = sample users from R
4        γ = constant
5        For user i in user_sample:
6            U = random sample from uniform distribution
7            Item_sample = sample items from user i's item rating list
8            Item_list = sorted item samples in decreasing rating values
9            V = random sample from uniform distribution
10           For j in 1: max_index_Item_list -1:
11               For k in j+1: max_index_Item_list:
12                   If R[i, j] > R[i, k]:
13                       U_i = U_i − γ ∂L/∂U_i
14                       U_w = U_w − γ ∂L/∂U_w
15                       V_j = V_j − γ ∂L/∂V_j
16                       V_k = V_k − γ ∂L/∂V_w
16   Reconstruct R by dot products of U and V
```

## V. EXPERIMENTS

We test Skellam Rank with 9 other modern day algorithms in our experiments. The hardware we use to test our algorithm is a Lenovo Laptop with Intel i5 2.5GHz CPU and 16GB RAM memory. The operating system on top of the laptop is Windows 11 operating system.

The 9 other algorithms used in our experiments are ZeroMat [33], Random Placement, Classic Matrix Factorization, DotMat [34], DotMat Hybrid [34], PoissonMat [35], PoissonMat Hybrid [35], ParaMat [36], and Pareto

Pairwise Ranking [30]. To save the time for discussion, we omit the introduction of the details of the algorithms.

We test our algorithms on 2 datasets: MovieLens 1 Million Dataset [37] and LDOS-CoMoDa Dataset [38]. MovieLens 1 Million Dataset comprises of 6040 users and 3706 movies with approximately 1 million user movie rating values. LDOS-CoMoDa dataset is a context-aware movie dataset that contains 121 users and 1232 movies with multiple contextual information fields.

The experimental results on MAE (Fig.1) lead to the conclusion that Skellam Rank is highly competitive with other algorithms, even when compared against modern day state of the art approach such as PoissonMat Hybrid and DotMat Hybrid, which are far superior to the classic matrix factorization algorithm.

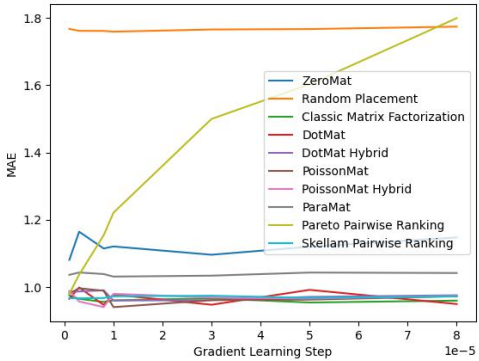

Fig. 1. Comparison of Skellam Rank against other modern day recommender systems on MAE score (*MovieLens 1 Million Dataset*).

We also test the algorithms on the same dataset with a fairness metric named Degree of Matthew Effect (DME) [5]. The comparison results are demonstrated in Fig.2:

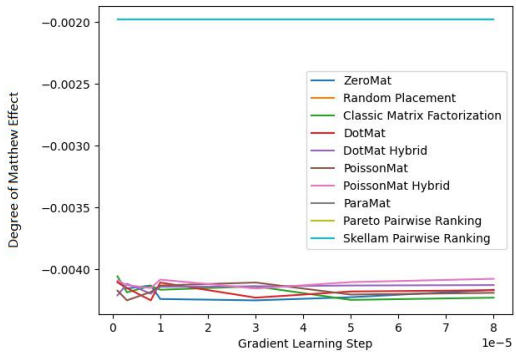

Fig. 2. Comparison of Skellam Rank against other modern day recommender systems on DME score (*MovieLens 1 Million Dataset*).

We also compare the algorithms on LDOS-CoMoDa dataset and evaluate the results on MovieLens 1 Million Dataset and LDOS-CoMoDa Dataset. The results could be found in Fig. 3 and Fig. 4:

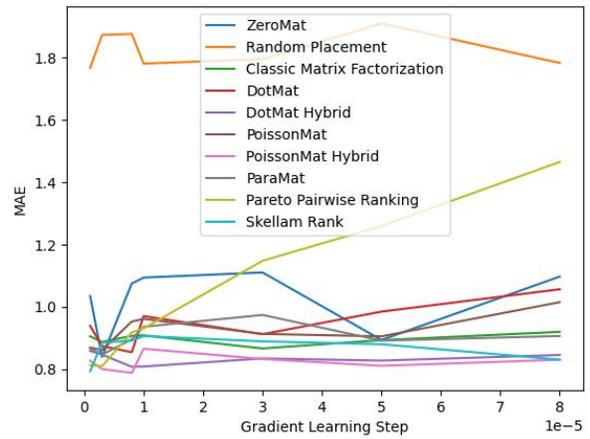

Fig. 3. Comparison of Skellam Rank against other modern day recommender systems on MAE score (*LDOS-CoMoDa Dataset*).

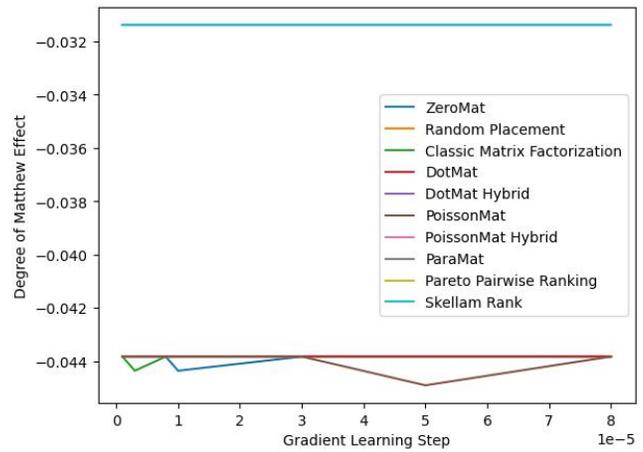

Fig. 4. Comparison of Skellam Rank against other modern day recommender systems on DME score (*LDOS-CoMoDa Dataset*).

## VI. CONCLUSION

In this paper, we introduced a new fair learning to rank algorithm named Skellam Rank. The algorithm is built upon the theory of Poisson process and Skellam distribution. We demonstrate in our experiments that our method is highly competitive among other algorithms when evaluated by accuracy metrics such as MAE score (accuracy), but is far superior to other algorithms when evaluated by DME score (fairness).

In future work, we would like to investigate into the explainability issue of the fair AI algorithms, especially when they are not built upon the regularization technique. We would also like to explore the solution to other problems related to fairness other than popularity bias.